\documentclass[
    ,final            
  ]
  {aipproc}

\layoutstyle{6x9}

\begin{document}

\title{Long Lived Charged Massive Particles and Big Bang Nucleosynthesis}

\classification{95.35.+d, 11.10.Kk, 12.60.-i}
\keywords      {Big Bang Nucleosynthesis, Long Lived Massice Charged Particles}

\author{Kazunori Kohri}{
address={Insitute for Theory and Computation, Harvard-Smithsonian Center for Astrophysics, 60 Garden Street, Cambridge, MA 02138, USA}}

\author{Fumihiro Takayama}{
  address={Institute for High Energy Phenomenology, Cornell University, Ithaca, NY 14853, USA}
}

\begin{abstract}
We consider Big Bang Nucleosynthesis(BBN) with long lived charged massive 
particles. Before decaying, the long lived massive particles recombines with 
a light element to form a bound state like a hydrogen atom. 
We discuss the possible change of primordial light element abundances due to 
formations of such bound states. 
\end{abstract}

\maketitle


\section{Introduction}
Under the remarkable triumph of cosmological standard model, 
the existence of dark matter has been well-established as one of the best 
evidence of new physics. 
Continuous efforts to identify what the dark matter is have
 been made in cosmological observations and collider experiments.
 In present stage, the only macroscopic phenomenon tell us the existence
 of dark matter, on the other hand, in future collider experiments and
 dark matter direct detection observations, the microscopic properties of 
dark matter may be understood, which has been one of the interesting targets 
for particle physicists.

Under several considerations on the extension of the standard model of 
elementary particle physics, a lot of candidates of dark matter has been
 proposed. In some scenarios as superWIMP dark matter~\cite{superWIMP}, 
it is not necessary that the dominant component of matter during deeply 
radiation dominated universe is neutral and the relic is same as present 
observed relic density of dark matter. Actually the extremely long lived 
CHarged Massive Particle(CHAMP) have been well-motivated , for example, 
in scenarios of gravitino LSP~\cite{gravitino} in the context of 
supersymmetric extension of particle standard model, and they may be
 the dominant matter in deeply radiation dominate era. 
Then the attractive candidate of such a long lived CHAMP may be stau NLSP 
which is a supersymmetric partner of tau lepton. 
If such long lived CHAMPs exist during BBN era, 
such a CHAMP may constitute a bound state with a light element and 
the formation of such bound states may modify nuclear reaction rates in BBN 
and eventually change light element abundances. 
The effects on light element abundances due to energy injections from 
late time decays have been performed assuming freely propagating decay 
particles~\cite{BBN}. On the other hand, in this paper, 
we consider new effects on BBN due to formation of such bound states which
 have not been discussed before~\cite{boundstate1}.
\section{CHAMP Big Bang Nucleosynthesis (CBBN)}
When the temperature becomes higher than the binding energy of the bound state,
 the photo-destruction rate of bound states is rapid. Then only small fraction
 of bound states can be formed. But once the temperature becomes lower than 
the binding energy, the formation of bound states become efficient.

\begin{table}
\begin{tabular}{|rc|rc|rc|} \hline
Nucleus(X) & & binding energy (MeV) & & atomic number & \\ \hline
$p$ & &0.025 & &Z=1 & \\
D & &0.050 & &Z=1 & \\
T & &0.075 & &Z=1 & \\
$^3$He & &0.270 & &Z=2 & \\
$^4$He & &0.311 & &Z=2 & \\ 
$^5$He & &0.431 & &Z=2 & \\
$^5$Li & &0.842 & &Z=3 & \\
$^6$Li & &0.914 & &Z=3 & \\
$^7$Li & &0.952 & &Z=3 & \\
$^7$Be & &1.490 & &Z=4 & \\
$^8$Be & &1.550 & &Z=4 & \\ 
$^{10}$B & &2.210 & &Z=5 & \\ \hline
\end{tabular}
\caption{Table of the binding energies for the various nuclei 
in the case of $Z_C=1$ given in Ref.~\cite{Cahn:1980ss}. 
For heavier elements than $^8$Be, 
the binding energies are given by the harmonic oscillator
approximation.}
\label{table:ebin}
\end{table}

By using the detailed balance relation between the forward process
 $X+C\to\gamma+(X,C)$ and the reverse process
$(X,C)+\gamma\to X+C$, and assuming SBBN processes are well decoupling, 
the Boltzmann equation for the capture reaction of a light element is,
\begin{eqnarray}
\lefteqn{ \frac{\partial}{\partial t}n_{X} +3Hn_{X}
} 
\nonumber \\
&\simeq& -\langle \sigma_r v \rangle \left[
  n_Cn_X-n_{(C,X)}n_{\gamma}(E>E_{bin})\right],
\end{eqnarray}
where
$n_{\gamma}(E>E_{bin})\equiv n_{\gamma}
\frac{\pi^2}{2\zeta(3)}(\frac{m_X}{2\pi T})^{3/2}
e^{-\frac{E_{bin}}{T}}$, $n_{\gamma}=\frac{2\zeta(3)}{\pi}T^3$, $T$ is thermal
 temperature, $E_{bin}$ is a binding energy and $n_i$ is the number density of
 $i$ species.
Then we can obtain the critical temperature that the formation of bound states
 dominate against the photo-destruction as follow
~\cite{boundstate1}.
\begin{eqnarray}
 T_c \simeq \frac{E_{bin}}{40}.
\end{eqnarray}
The binding energies for each bound states are shown in Table~\ref{table:ebin}.
 We can find that the bound state formation for heavier elements may occur 
in earlier time. Such bound state may cause some modification of 
nuclear reaction rate in BBN and cause changes on light element abundances.
 
Next we investigate the effects on BBN due to bound state formations.
We consider the thermal freeze out of light element abundances in CBBN
and here we simply ignore the effects of possible high energy
injections due to the late decay of CHAMPs,  which may provide the
initial condition to consider such late decay phenomenon if the decay
occurs enough after the decoupling of  the BBN processes. 
In our estimation, we also assume the instantaneous captures
for each light elements at $T_c=E_{bin}/40$.  

\begin{figure}
  \includegraphics[height=.3\textheight]{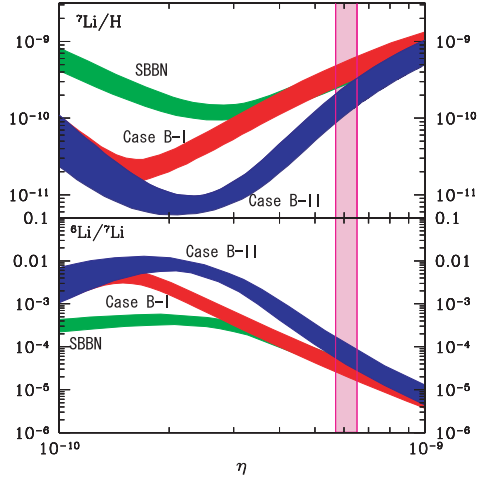}
  \caption{Prediction of $^7$Li/H ({\it upper panel}) and 
$^6$Li/$^7$Li ({\it lower  panel}) as a function of $\eta$. 
The red (blue) band is for Case B-I (Case B-II) in
CBBN. Here we assumed $n_{C}/n_{\gamma}=3.0\times 10^{-11}$ and
the instantaneous capture of CHAMPs. The Case B--I means that
$E_{CM}=(\mu_{ab} /\mu_{(aC)b})E_{bin}+E_0$ in a process
$(a,C)+b\to (c,C)+d$ where we take $E_0$ to be  the Gamow's peak
energy for collisions between two charged elements, and to be $3T/2$
for collisions between a nucleus and a neutron. The Case B-II means
that we take $E_{CM}=E_{bin} +E_0$ as the CM energy of
processes and ten times larger value of the p-wave part of the cross
section of $^7$Be(n,$\alpha$)$^4$He than that in the standard BBN
code. It is showed that the modification
by a factor of ten on the p-wave partial cross section of
$^7$Be(n,$\alpha$)$^4$He does not change the  SBBN prediction (Case
B--I) but must be important in CBBN (Case B--II).}
\end{figure}

We evaluated the modifications of nuclear reaction rate from the SBBN values
~\cite{boundstate1}. 
 Once bound states were formed, an incident charged nucleus can penetrate
 the weakened Coulomb field and collide with the bound nucleus relatively 
rapidly. Also in SBBN, the kinetic energy of light elements is governed by 
thermal temperature, but after formation of bound states, 
the kinetic energy of bound light elements is the binding energy of bound 
states, which does not depend on 
 thermal temperature and even at low temperature, the nuclear reaction 
may be proceed with relatively large CM energy.

In SBBN, abundances of all light elements are
completely frozen until $T\sim$ 30keV. Since  $T_c$ is 24keV for
$^7$Li and 38keV for $^8$Be, which is almost the end  of  SBBN, 
the formations of bound states may change their abundances. On the other hand, 
for lighter elements than $^6$Li, since the efficient captures occur only
at below 10 keV,  the change of nuclear reactions can
not recover the processes  at such a low temperature.  This conclusion
will be hold if the difference from our estimation of nuclear reaction rate 
is not significant.\footnote{In \cite{boundstate2}, bound state effects that 
we have not considered here have been examined. } 

In Fig.1, we plot the theoretical
prediction of $^7$Li/H ({\it upper panel}) and $^6$Li/$^7$Li ({\it
lower  panel}) as a function of $\eta$. The SBBN predictions are plot
by the green bands. 

\section{Conclusion}
We discussed the effects on BBN due to the formation of bound states with a 
CHAMP. Such extremely long lived CHAMPs can be trapped and the
 detail properties can be examined in near future collider experiments
~\cite{Trap}. Also some considerations on CHAMP catalysis fusion~\cite{fusion} 
may be interesting to understand the effect on nuclear reactions due to 
bound state formations. 


\begin{theacknowledgments}
F.T would like to thank Bryan T. Smith, Jonathan L. Feng, Jose 
A.R. Cembranos for discussions in the early stage of this work and
Maxim Pospelov, Manoj Kaplinghat, Arvind Rajaraman for discussions in
SUSY06 and Mikhail Shifman, Keith Olive for suggestions during my visit 
to their institute and Koichi Hamaguchi, Tsutomu Yanagida for discussions 
on stau catalysis fusion and Toichiro Kinoshita, Maxim Perelstein for 
valuable suggestions. This work was supported in part by NASA grant NNG04GL38G.
\end{theacknowledgments}


\end{document}